# IoT Equipped Intelligent Distributed Framework for Smart Healthcare Systems

*Sita Rani[1], Meetali Chauhan[2], Aman Kataria[3] and Alex Khang[4]

[1]*Department of Computer Science and Engineering,*
*Gulzar Institute of Engineering and Technology,*
*( Affiliated To I.K.G. Punjab Technical University, Kapurthala)*
*Gulzar Group of Institutions, Khanna ( Ludhiana)*
*Punjab, INDIA- 141401.*
sitasaini80@gmail.com

[2]*Department of Computer Science and Engineering,*
*Gulzar Institute of Engineering and Technology,*
*( Affiliated To I.K.G. Punjab Technical University, Kapurthala)*
*Gulzar Group of Institutions, Khanna ( Ludhiana)*
*Punjab, INDIA- 141401.*
meetalichauhan08@gmail.com

[3]*Project Associate*
*CSIR-CSIO, Chandigarh*
*INDIA-160030*
ammankataria@gmail.com

[4]*Professor of Information Technology,*
*GRITEx and VUST,*
*VIETNAM .*
alex.khang@outlook.com

**Abstract:** The fundamental aim of the healthcare sector is to incorporate different technologies to observe and keep a track of the various clinical parameters of the patients in day to day life. Distant patient observation applications are becoming popular as economical healthcare services are facilitated by these apps. The process of data management gathered through these applications also require due attention. Although cloud facilitated healthcare applications cater a variety of solutions to store patients record and deliver the required data as per need of all the stakeholders but are affected by security issues, more response time and affecting the continues availability of the system. To overcome these challenges, an intelligent IoT based distributed framework to deploy remote healthcare services is proposed in this chapter. In the proposed model, various entities of the system are interconnected using IoTs and Distributed Database Management Systems is used to cater secure and fast data availability to the patients and health care workers. The concept of Blockchain is used to ensure the security of the patient's medical records. The proposed model will comprise of intelligent analysis of the clinical records fetched from Distributed Database Management Systems secured with Blockchain. Proposed model is tested with true clinical data and results are discussed in detail.

## 1. INTRODUCTION

Health is the most essential aspect of living. In current times, the modern society is suffering from a number of problems like multiple organs failure & various chronical diseases due to stress & anxiety. So, the society requires appropriate facilities, resources, and services from the hospitals such as medications, doctors & nurses on time [1, 2]. With sharp rise in chronic diseases & pandemic era of Covid-19, a sudden hike can be seen in usage of smart healthcare systems. In order to provide efficient healthcare services to the patients, a major role is played by smart healthcare system. This has reduced mandatory physical presence of patients at hospitals [3-5]. E-healthcare system have provided high quality care to the patients by providing online medical services at homes itself. Earlier, there was a communication gap between patients & doctors due to unavailability of doctors in case of emergency. But now, advanced communication systems and Internet of Things (IoT) technology has made this possible by providing effective communication paradigm. IoT is an appropriate solution to administer such problems occurring in healthcare systems. This paradigm is used to collect patients' data which is analyzed by doctors to provide medication & medical treatment remotely. This automation in smart healthcare monitoring system has reduced risk of patient's life by providing on time medical help to them in case of emergency. There are various monitoring systems such as sensors to collect data, IoT gateway to distribute data, and cloud-based storage for storing patients' data to get examined by doctors. IoT acts as a chain, responsible to collect all the information communicated from smart devices via internet. Patients can receive their health records using mobile healthcare apps [6].

**1.1 Internet of Things (IoT)**

IoT is a concept which reflects connectivity of devices, services, and system in terms of machine- to -machine and man- to –machine [7, 8]. This helps to achieve automation which can be seen in a wide range of areas, such as smart cities, healthcare, traffic management, logistics, and waste management. IoT has provided incredible outlook to modern healthcare system by facilitating human life with healthcare apps, fitness programs, remote health monitoring, emergency help, etc. In collaboration with medical devices, such as sensors & imaging devices, the service providers can provide best guidance to the patients. Thus, IoT based healthcare services are expected to reduce medical costs and increase for the patients.

Recently evolved IoT-based wireless technologies have helped to prevent & diagnose chronical diseases and provide real-time monitoring. Database records and servers play a vital role to maintain medical records, and provide facilities to the patients in need of hour. Table.1 describes the IoT integrated advanced technologies which are useful in domain of healthcare [9].

Table 1. IoT-integrated Technologies and Their Benefits in the Domain of Healthcare

| Technology | Description |
|---|---|
| Big data | Data is stored which provides a quick review easily in healthcare systems when required |
| | Helps to maintain clinical records, bills, medical history of patients |
| Cloud computing | Useful to store on demand data & specifies the content via internet |
| | This helps to visualize the data resources which helps the doctors to work effectively |
| Smart sensors | This device provides accurate results & monitors all medical parameters |
| | This device controls various aspects such as blood pressure, oxygen levels, temperature & sugar levels |
| Software | It is used to get associated with patients' data, medical tests & reports |
| | It reduces the communication gap between doctors & patients |
| Artificial intelligence | This concept evaluates, predict, analyze & helps in decision making |

|  | This concept with help of algorithms predict & controls diseases |
|---|---|
| Actuators | This device helps to maintain accuracy in calculated parameters |
|  | This device helps to control system & makes it act according to the requirements |
| Virtual Reality | It provides digital information & improves patients safety |
|  | It provides real-time information with integration of humans with electronic systems |

A generalized IoT- based automatic framework developed for healthcare systems is shown in Fig. 1. It describes the results predicted using IoT integrated technologies [10-12].

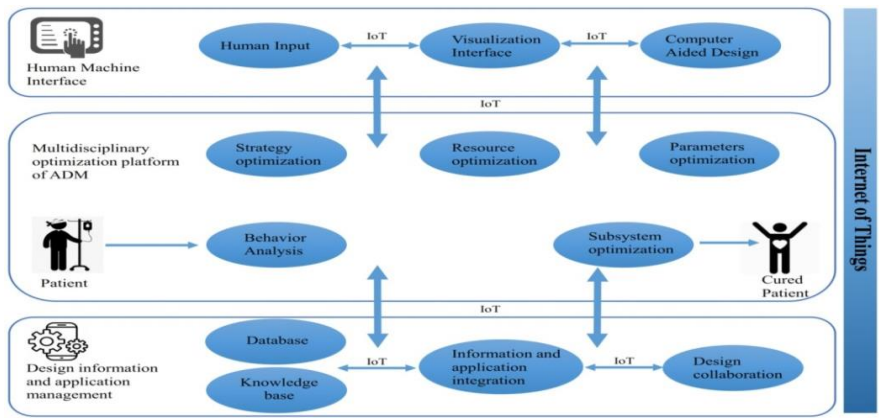

Fig.1 IoT- based Automatic Design Framework

The major role played by IoT in the domain of healthcare is in the areas of silent symptoms of patients. Early diagnosis might prevent severe illness & saves patients from untimely death. So, early diagnosis might save patients life. In current times, the IoT based health applications focus on medical treatments of the diseases and monitoring the health of the patients via analyzing various parameters using smart devices. The healthcare system is slowly switching to remote healthcare by providing E-health services at homes. Fig.2 shows the IoT-based healthcare application framework to facilitate the residents. There are many applications existing for patient monitoring. In addition, various networks such as Wireless Body Area Network (WBAN), Wireless Local Area Network (WLAN), Radio-frequency Identification (RFID) and Wireless Personal Area Network (WPAN) assist in automatic identification and data capturing [13].

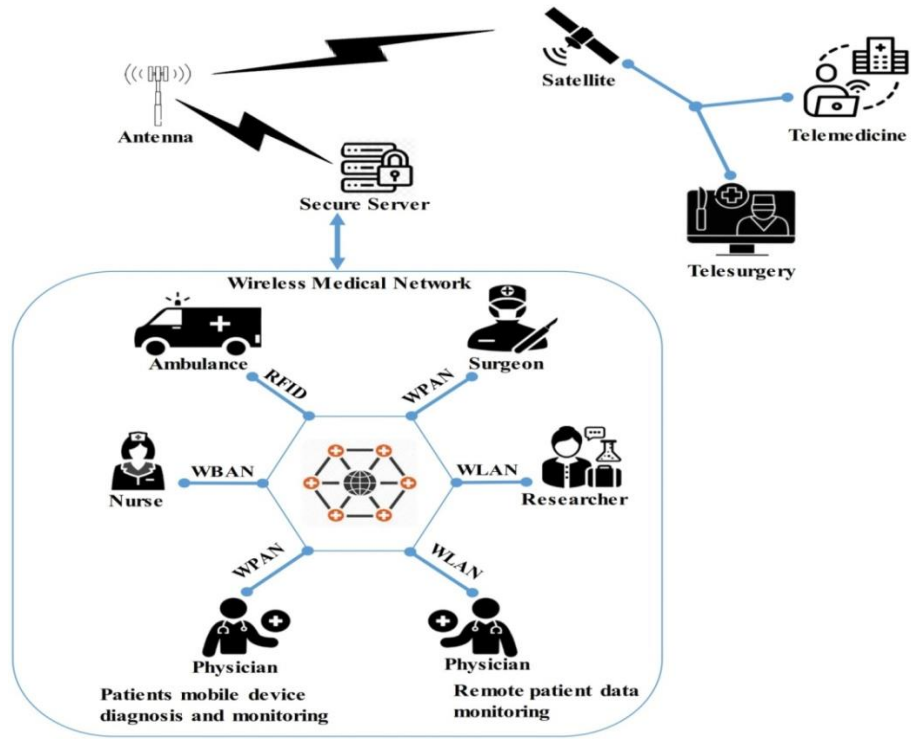

Fig.2 IoT-based Healthcare Application Framework

**1.2 Smart Healthcare**

Using traditional healthcare facilities, it was difficult to diagnose and treat patients in case of emergency, which was used to cause mental trauma, cardiovascular disorders, anxiety & depression to the patients and their family members. With launch of smart healthcare applications and services a wide variety of online facilities got available to people at homes.

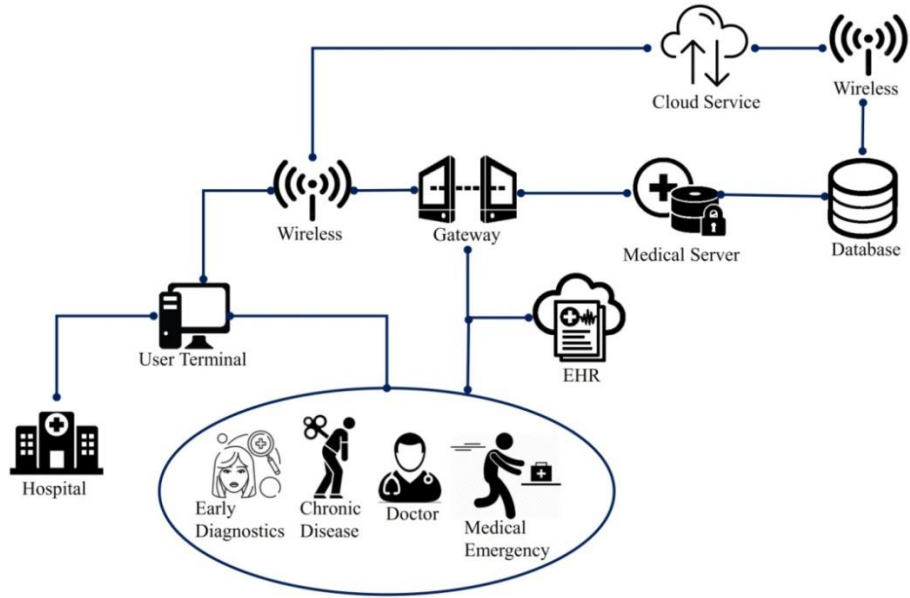

Fig.3 Smart Healthcare Services

Fig.3 highlights some of the smart healthcare services which have provided ease to human beings with secure services, such as availability of online appointments with doctors, storing medical records of patients, and consultancy to the patients in case of emergency [14]. The whole healthcare system is connected via wireless technology and database is maintained using cloud computing.

Table 2 shows various smart applications launched with the purpose to make them available all-time to the people. These applications facilitate the people with various health services. They can analyze various parameters themselves on daily basis and opt workout & healthy eating habits accordingly. Various remedies can be followed without consulting doctors just by adapting healthy lifestyle [9, 15].

Table 2. Healthcare Applications and Services

| Application | Services |
|---|---|
| Health Assistant | Body temperature, fat, weight, BP, glucose level check |
| Calorie Counter | Calories count from the food eaten |
| Pedometer | Steps taken and calories burnt |
| Period Tracker | Record of menstrual cycle in women |
| Google Fit | Running, cycling and walking activities |
| Water Your Body | Water drinking habits and alerts |
| Heart Rate Monitor | Heart beat |
| Smart Watch | Number of steps taken, BP, heart rate, calories burnt |
| On Track Diabetes | Blood glucose level |
| Finger Print Thermometer | Body temperature |

## 1.3 DBMS

IoT based distributed technology plays a vital role in smart healthcare system. Distributed healthcare system has interconnected various medical resources with sensors and actuators such as ECG, BP machine, EMG, and Glucometer. With sharp rise in usage of IoT, there has been a major contribution of distributed database management system (DDBMS) to provide efficient healthcare services to patients. The distributed healthcare system acknowledges various parameters such as blood pressure (BP), sugar levels of diabetic patients etc. via health monitoring devices. Efficient remedies & medical prescription can be suggested by doctors in case of emergency to the patients. The purpose of DDBMS is to extend the healthcare services from various domains of hospitals to patients at homes. The DDBMS in collaboration with IoT-based healthcare are dedicated to provide various services using a variety of healthcare applications with other important resources which include laptop, mobile phones, sensors, actuators, medical equipment, e-patients, patients records stored using cloud computing, medical staff of doctors & nurses, and database of patient's records. Fig.4 shows collaborative healthcare system comprising IoT & DDBMS [16].

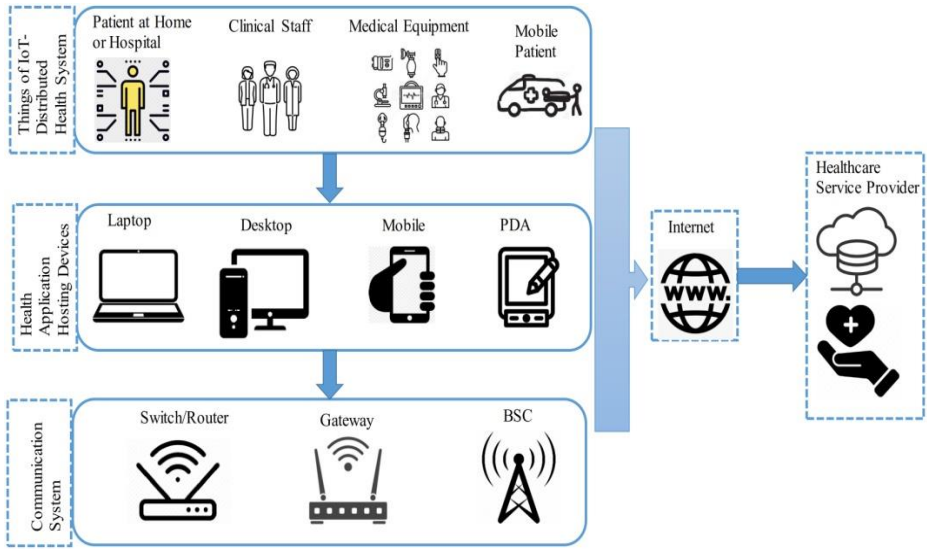

Fig.4 Distributed Healthcare System

**1.4 Artificial Intelligence (AI)**

Artificial Intelligence (AI) is a concept which has no boundaries in terms of smart development. It contains the ability of algorithms in machines to analyze the results without human intervention. The induction of AI with smart devices in the domain of healthcare has set an ideology for healthcare systems to a new level [17]. With advancement in healthcare systems, health status of people has reached a new level. AI embedded machines comprising sensors help to monitor and diagnose symptoms of diseases at earlier stages. This whole system acts like a robotic nurse for patients who takes care, monitor, and record patient's health condition with consistency. Using AI algorithms, various roles & responsibilities are carried out which requires intelligence in the areas of image analysis, speech recognition, under water communication, pattern recognition & decision making [18-23]. This provides assistance to suggest best way to cure health. The collaboration of AI algorithms makes it easier to make accurate predictions in laboratories, such as blood group detection, and disease predictions [24]. So, AI is the preferable choice when it comes to decision making [25].

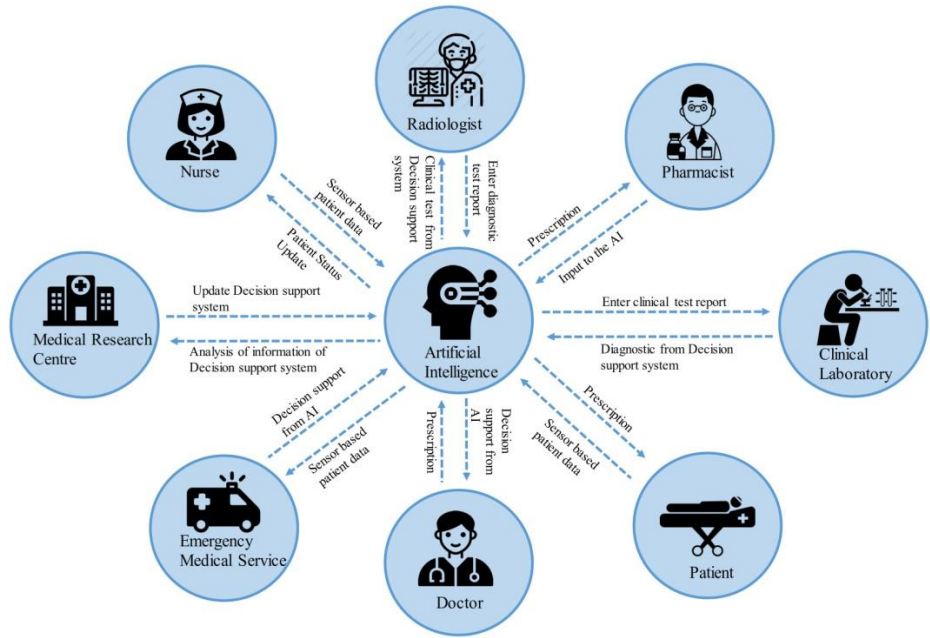

Fig.5 Smart Healthcare using AI

AI algorithm is a mixed composition of various technologies which includes natural language processing (NLP), machine learning, neural networks & robotic systems. AI is used in multiple fields of smart healthcare, such as cancer treatment surgeries, neurology, and cardiology. Fig.5 provides an outlook of AI based healthcare system. AI based healthcare system requires a balanced approach which can be achieved through support from all domains such as doctors, nurses, labs, radiologists, pharmacy, and emergency medical services.

**1.5 Blockchain Technology**

Blockchain in healthcare is linking of patient medical records, doctors, hospitals, nurses, medical staff , and health communities for the welfare of patients [26].

Blockchain is a kind of framework used to provide secure data exchange and management process [27]. The basic idea of blockchain is to share data via peer-to-peer network. This helps to communicate data to authenticated users. They can modify or delete data records accordingly. In healthcare sector, the presence of sensitive information which needs to be secured from third party users to maintain a privacy & security is critical. Due to such sensitive content, blockchain concept has been incorporated in healthcare sector to handle security systems. In addition, the blockchain concept has been used widely to resolve problems of central administration in database securely [28] **.**

Blockchain has eliminated the need to govern or manage the authentication based on trust & transparency. Blockchain has further enhanced in terms of privacy & security using cryptographic hash functions. Some of the blockchain applications in healthcare are compilation of visitor's details, patients' records, records of lab results, and treatment details of patients. All such details are ensured by blockchain process which includes ambulatory services & data assistance. A commonly observed problem with medical data is the duplication or mismatch of details during analysis of patient's records. Such issues are tackled using hash function in the blockchain process which includes hashed ledger instead of using a primary key. In addition to this, a blockchain concept is based on certification, due to which claims can be automatically verified whenever required. Blockchain has reduced data compilation as well as cost [29]. It has also reduced wastage & chances of fraud due to digitization of complicated datasets [30].

## 2. SECURITY ISSUES IN SMART HEALTHCARE SYSTEMS

There are many security issues in the deployment of smart healthcare systems, discussed below:

### 2.1 Communication Media

Healthcare devices are connected to global as well as local networks via wide range of wireless links such as Bluetooth, GSM, WIFI, and Zigbee. But wireless network makes traditional security schemes less appropriate. Therefore, it is very difficult to manage security protocols which can handle both wired and wireless technologies equally [31].

### 2.2 Topology Issues

IoT-integrated healthcare devices are connected over various types of the network for data collection, storage, and computation [32]. But the problem occurs when they exit from network due to certain failure. This cause dynamic network topology issues [33, 34].

## 2.3 Scalability

With increase in IoT devices, there is requirement to integrate them in global information network. Therefore, to design a scalable security scheme without considering security related requirements become a tough task [6, 35].

## 2.4 Mobility & Energy Constraints

IoT based devices are dynamic in nature. They work on batteries. But, as different networks have different configurations, more efficient security algorithms are required for mobility [36, 37].

## 2.5 Memory Constraints

Most of the IoT based devices have limited memory, which is one of the major issues faced in storage of data and functioning of the devices [38].

## 2.6 Multi-Protocol Network

IoT devices communicate with other devices over the local network using network protocols. In addition to this, some IoT devices communicate with IoT service providers via IP network. But security experts face problems related to sound security solutions for multiple protocol communication [39].

**2.7 Tamper Devices**

An attacker might try to tamper the devices in physical mode and may extract cryptographic secret content. He might modify, delete, or replace the content with malicious data [6, 40].

**3. EXISTING HEALTHCARE SYSTEMS**

The task of data management plays a very important role to administer different types of the services in smart healthcare systems. The efficient storage of patient health data facilitates disease diagnosis, vaccine scheduling, and deployment of other health services. In the current scenario of smart healthcare framework, penitent health data is maintained in the electronic form using cloud architecture. Over the time, different authors have proposed different secure techniques and frameworks to store patients' data. In [41], authors presented a novel approach for sharing patient data in a secure way. Using this technique, patients' data is organized and shared through a semi-trusted server. In this technique, every attribute of each patient record is saved and transmitted in encrypted form. The scalability of the medical data is the main feature of this method. In [42], another conflux approach using encryption mechanism and digital signature is proposed for secure transfer of medical records. In [43], authors discussed the disadvantages of the approach presented in [42], and introduced a new method to overcome the issues. Authors in [44-46], proposed a number of authentication mechanisms and data transfer protocols to secure the storage and transfer of medicals records on different types of machines and mobile devices. Many authors proposed different privacy preserving techniques to secure electronic medical records in the distributed smart healthcare systems. Although, lot many efforts are put by the researchers to secure medical data of the patients in smart medical systems adopted by various hospitals, this can also make the faster access of the records inconvenient in emergency situations. Emergency care providers and doctors may face hurdles to provide first aid and other medicals services. To resolve these challenges, medical industry, researchers, and academicians have introduced many smart gadgets/devices to monitor individuals' health and store health parameters [47, 48]. But these devices are vulnerable to data thefts and failures [49].

To address various security issues in exiting healthcare systems, many blockchain models are proposed by the researchers. More secure solutions are proposed by storing the hash tables for cloud data in the blockchain nodes [50-53]. Another, more secure technique is presented by the authors in [54]. This model is proposed to access medical records by the doctors and patients. In [55], authors proposed a multi-workflow-based system to manage various processes, like clinical trials and complicated surgeries. To store patient's medical data more securely, and manage personal information efficiently, a novel platform framework is proposed by the authors [56]. Remote monitoring of the patients and wearable gadgets are highly prone to data stealing [2]. To identify malfunctioning devices in a network is also a tedious job. Many authors carried out their work focusing these challenges. In [57], authors proposed a novel model to address the privacy challenges in remote monitoring systems. Few other systems are also proposed by the researchers for secure smart-contract based remote patient monitoring [28, 58, 59].

Table 3. Comparative Analysis of the Exiting Work with Proposed Work

| Ref. No. | Year | IoT-Facilitated Remote Monitoring | Medical Records | AI-Based Smart Contract | Distributed Record Storage | Locate Malfunctioning IoT-Device |
|---|---|---|---|---|---|---|
| [55] | 2020 | ✘ | ✓ | ✓ | ✓ | ✘ |
| [57] | 2018 | ✓ | ✓ | ✓ | ✓ | ✘ |
| [58] | 2019 | ✓ | ✓ | ✓ | ✓ | ✘ |
| [28] | 2020 | ✘ | ✓ | ✓ | ✓ | ✘ |
| [59] | 2020 | ✘ | ✓ | ✓ | ✓ | ✘ |
| [60] | 2020 | ✘ | ✓ | ✓ | ✓ | ✘ |
| [61] | 2020 | ✘ | ✓ | ✓ | ✓ | ✘ |
| Proposed Work | | ✓ | ✓ | ✓ | ✓ | ✓ |

The model presented in this paper stores patient health records in a blockchain based network. This secure system is proposed for IoT-based patient health monitoring. Smart contract is supported by Artificial

Intelligence to locate the malfunctioning nodes. We have used 5 different attributes to characterize the performance of the proposed model. Main features of the various existing solutions proposed by the authors, and their comparison to the proposed system is summarized in Table 3.

## 4. PROPOSED MODEL

The proposed framework is implemented in four layers, i.e., Hospital, remote IoT-integrated medical nodes, distributed medical records, and AI-based smart contract, as shown in Fig. 6.

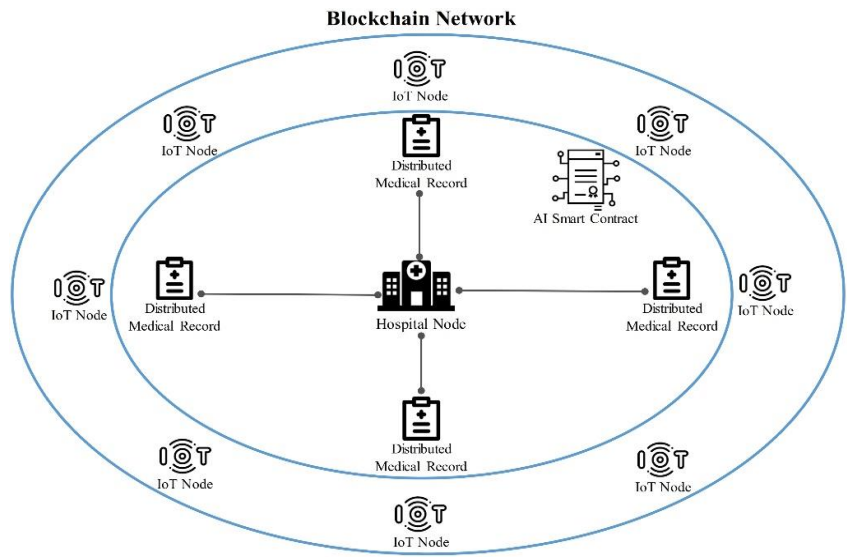

Fig. 6 Proposed Model: Framework

- **Hospital-** Basically, this layer acts as an information warehouse. It keeps complete patients' information using various attributes;

few important ones are patient id, patient name, disease history, and medicines prescribed.

- **Distributed Medical Records-** This layer is integrated with hospital layer. Medical records stored in the hospital layer are distributed across the different nodes to make them secure using blockchain network.
- **AI-based Smart Contract-** This layer is merged between hospital and distributed medical records layer, IoT nodes and hospital layer, and IoT nodes and distributed medical records layer. It aids the process of decision making, breach detection, and to check for malicious data.
- **Remote IoT-integrated Medical Nodes-** These nodes are used to sense the various health parameters of the patients and transfer the sensed data securely to blockchain protected distributed databases.

Working of the proposed model is depicted with the flowchart shown in Fig. 7.

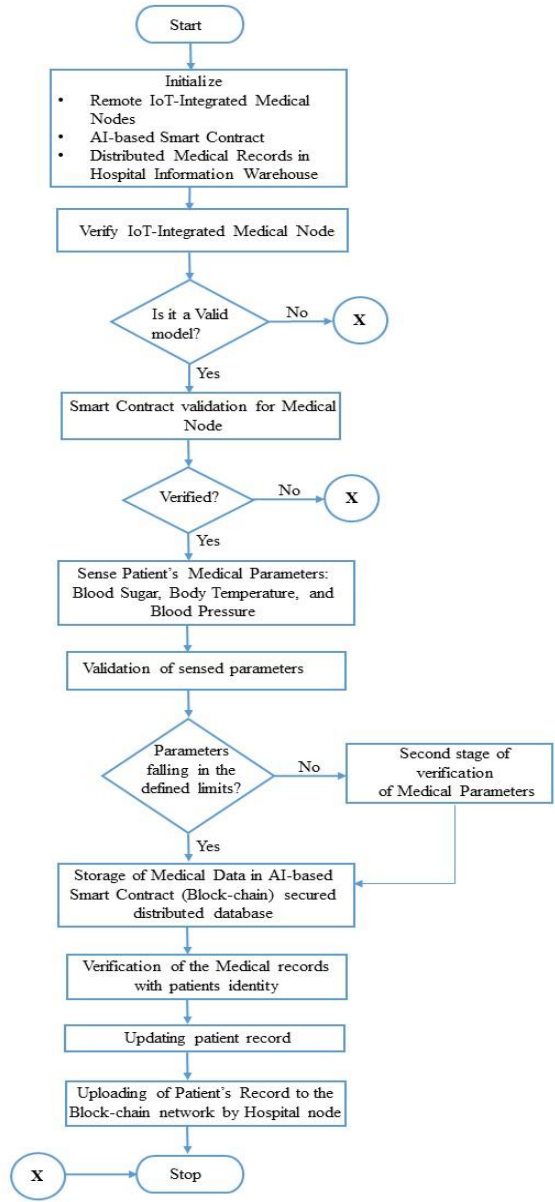

Fig. 7 Proposed Model: Working

## 5. RESULTS AND DISCUSSION

This section focusses on the results obtained with the proposed model. The proposed model is evaluated using time taken by the transaction, throughput, and latency. To gather results, total 5 different nodes were deployed, i.e., 4 IoT devices to sense data, and one hospital node. The blockchain to secure distributed medical records on the hospital node was deployed using the Ethereum platform. Our work is deployed on the blockchain using AI supported smart contract. The results obtained for the deployed IoT devices, i.e., transaction processing time and average delay are shown in Table 4 and Table 5 respectively.

Table 4. Transaction Processing Time for IoT-based Medical Devices

| Number of Transactions | Processing Time (Seconds) | | | |
|---|---|---|---|---|
| | Device 1 | Device 2 | Device 3 | Device 4 |
| 50 | 20 | 22 | 18 | 23 |
| 100 | 31 | 41 | 37 | 45 |
| 150 | 55 | 62 | 58 | 67 |
| 200 | 82 | 84 | 73 | 91 |

Table 5. Average Delay in Transaction Processing for IoT-based Medical Nodes

| Number of Transactions | Average Delay (Seconds) | | | |
|---|---|---|---|---|
| | Device 1 | Device 2 | Device 3 | Device 4 |
| 50 | 0.8 | 0.9 | 0.7 | 0.8 |
| 100 | 2.1 | 2 | 1.9 | 1.7 |
| 150 | 2.4 | 1.8 | 2.6 | 2.1 |
| 200 | 3.6 | 3.2 | 3.3 | 3 |

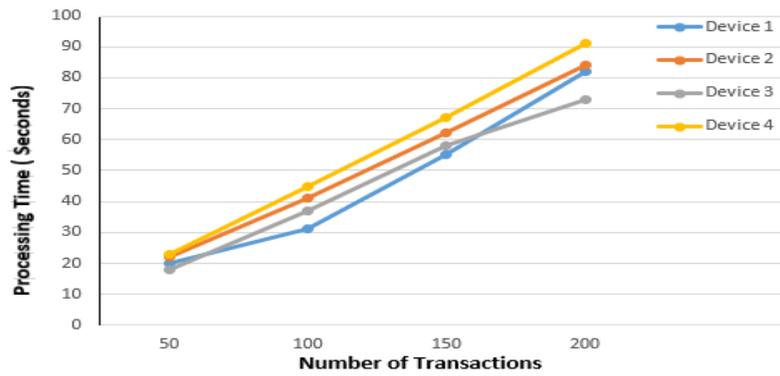

Fig. 8 Transaction Processing Time: Medical IoT Nodes

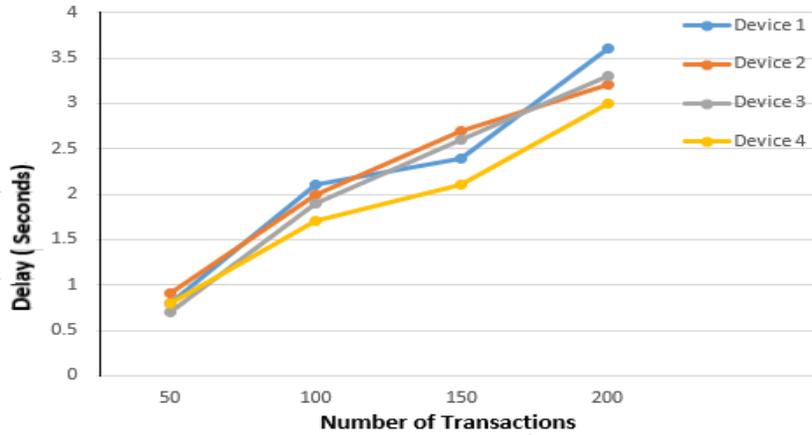

Fig. 9 Average Delay: Medical IoT Nodes

As discussed above, two important parameters considered to evaluate the deployed model are transaction processing time and average delay. As shown in Fig.8 and Fig.9, transaction processing time and average delay for all used IoT-based medical devices is almost similar. So, the proposed model is highly scalable in terms of number of smart medical devices as well as number of transactions executed. Our proposed model ensures blockchain based secure storage and access of distributed medical records in a smart hospital framework.

## 6. CONCLUSIONS

Smart distributed healthcare systems are rapidly becoming popular. They are fulfilling the medical needs of modern society in more transparent, secure and convenient way. Along with all these features, the proposed blockchain model also protects the healthcare system from a single point of failure using AI-based smart contract technique deployed at the hospital layer. The medical records of the patients are stored in distributed databases. The proposed framework is tested in real time environment for two important parameters, i.e., transaction processing time and average delay. The proposed system can further be enhanced by incorporating more efficient AI algorithms to improve the processing time in the blockchain network to minimize average delays.